\documentclass[aps,twocolumn,amsmath,amssymb,prx,superscriptaddress,preprintnumbers,nofootinbib]{revtex4}
\usepackage{graphicx}
\usepackage{color}

\newcommand{\bfl}{S}
\newcommand{\bfll}{T}
\newcommand{\sga}{i}
\newcommand{\sgaa}{j}
  
 \usepackage{slashed}
\usepackage{tikz}
\begin{document}
\title{Non-Invertible Chiral Symmetry and Exponential Hierarchies}
\author{Clay C\'{o}rdova}
\affiliation{Enrico Fermi Institute and Kadanoff Center for Theoretical Physics, University of Chicago}
\author{Kantaro Ohmori}
\affiliation{Faculty of Science, University of Tokyo}

\begin{abstract}
\noindent We elucidate the fate of classical symmetries which suffer from abelian Adler-Bell-Jackiw anomalies.  Instead of being completely destroyed, these symmetries survive as non-invertible topological global symmetry defects with worldvolume anyon degrees of freedom that couple to the bulk through a magnetic one-form global symmetry as in the fractional hall effect. These non-invertible chiral symmetries imply selection rules on correlation functions and arise in familiar models of massless quantum electrodynamics and models of axions (as well as their non-abelian generalizations).
When the associated bulk magnetic one-form symmetry is broken by the propagation of dynamical magnetic monopoles, the selection rules of the non-invertible chiral symmetry defects are violated non-perturbatively. This leads to technically natural exponential hierarchies in axion potentials and fermion masses.
 
\end{abstract}
\maketitle

\section{Introduction}\label{intro}

Symmetry in its myriad incarnations is a unifying principle of quantum field theory.  It organizes universality classes, and provides one of the few generally applicable tools to constrain correlation functions. Over the past few years, the concept of symmetry has undergone a profound sequence of generalizations broadening its scope and applicability. In this paper, we continue these developments by exhibiting new symmetries in familiar models of massless quantum electrodynamics and models of axions (as well as non-abelian generalizations).  We further explore mechanisms for weakly breaking these novel symmetries and illustrate that this often leads to technically natural exponential hierarchies in effective field theory.

\subsection{The Expanding Paradigm of Global Symmetry}

The core idea behind recent progress is the intrinsic formulation of internal symmetry in quantum field theory by topological operators \cite{Gaiotto:2014kfa}.  In this point of view, an ordinary global symmetry gives rise to a topological operator of codimension-one (equivalently the dimension of space).  These symmetry defects act on local operators and impose selection rules on correlation functions.  The fact that they are topological, i.e.\ invariant under deformations of their support that do not intersect other operators, encodes in an abstract geometric fashion the fact that these symmetries are conserved.  Meanwhile, the fact that ordinary symmetries form a group is captured by the symmetry defect fusion algebra: the operator product of defects associated to group elements $g_{1}$ and $g_{2}$ leads to the defect associated to $g_{1}g_{2}$.

From this starting point there are two significant generalizations that appear in an interlocked manner below:

\emph{Higher-Form Global Symmetry}  \cite{Gaiotto:2014kfa}. In this generalization, the topological defects defining the symmetry are allowed to have general codimension-$(q+1)$, corresponding to a so-called $q$-form global symmetry.  Here the charged objects are extended operators of dimension $q$.  Most relevant for our discussion is the example $q=1$ in four-dimensional quantum field theory.  In this case, the associated symmetry defects are topological surface operators.  These operators frequently arise in gauge theory, and provide a symmetry-based view of confinement.

A particularly simple example of such a symmetry which will feature prominently below is the $U(1)^{(1)}$ magnetic one-form symmetry of abelian gauge theory.  Denoting the dynamical field strength by $f$, this symmetry is generated by the current $*f$ which is tautologically conserved.  The associated charged line defects are 't Hooft lines which physically model infinitely massive magnetic monopoles.

\emph{Non-Invertible Global Symmetry}. In this generalization, the topological defects defining the symmetry are permitted to have a more general fusion product, beyond that captured by a group.  For instance, the fusion of two such defects may involve a sum of other defects, representing a variety of possible fusion channels.  This algebraic structure is referred to as \emph{non-invertible} to emphasize that, in contrast to group-like symmetries, a typical topological defect need not admit an inverse under fusion.

A variety of examples of this rich structure have recently been constructed.  In two spacetime dimensions, the fusion categories of lines are well-known from rational conformal field theory \cite{Fuchs:2002cm, Frohlich:2004ef, Frohlich:2006ch} and have been investigated in \cite{Bhardwaj:2017xup, Chang:2018iay, Thorngren:2019iar, Thorngren:2021yso}, and applied to constrain the dynamics of gauge theories \cite{Komargodski:2020mxz}.   In spacetime dimension larger than two such defects have also recently been constructed by generalizing Kramers-Wannier duality to higher spacetime dimensions \cite{Koide:2021zxj, Choi:2021kmx, Kaidi:2021xfk, Hayashi:2022fkw, Choi:2022zal, Kaidi:2022uux}, as well as by novel notions of gauging \cite{Roumpedakis:2022aik}.   Their algebraic properties have also been developed in \cite{Roumpedakis:2022aik, Bhardwaj:2022yxj}.

The dynamical implications of these generalized symmetries are yet to be fully understood.  This presents an opportunity for investigation that sets the context of this work.

\subsection{Non-Invertible Chiral Symmetry Defects}

Let us now focus on the main class of examples described below.  Our first result, derived in section \ref{secABJ}, is to clarify the nature of chiral symmetry in abelian gauge theory.  In general, a common understanding is that a classical $U(1)_{\text{chiral}}$ symmetry with an ABJ anomaly \cite{Adler:1969gk, Bell:1969ts} is destroyed quantum mechanically.  

Instead, as we argue below,  in a certain cases the chiral symmetry is not destroyed, but its nature is changed from an ordinary invertible symmetry classically, to a non-invertible symmetry quantum-mechanically.  One such case is when the gauge symmetry that causes the ABJ anomaly is abelian, which we call an abelian ABJ anomaly. More generally, this can also happen when the gauge group has a non-trivial fundamental group.

In more detail, we construct non-invertible topological defects $\mathcal{D}_{k}$ which act on local operators as a discrete chiral symmetry rotation by a $k$-th root of unity.  
Semi-classically, these defects may be understood as a composite of the naive invertible chiral symmetry defect $\mathcal{C}_{k}$ fused with a three-dimensional topological field theory whose spectrum of lines is precisely chosen to compensate for the anomaly of $\mathcal{C}_{k}$.  In the terminology of \cite{Hsin:2018vcg}, the relevant topological theory is a minimal abelian TQFT $\mathcal{A}^{N,p}$, where the index $N$ specifies that the theory has $\mathbb{Z}_{N}^{(1)}$ one-form symmetry, and $p$ controls the spins of the associated abelian anyons (see \eqref{spinp} below). In equations, we thus write
\begin{equation}
\mathcal{D}_{k}\equiv \mathcal{C}_{k}\times \mathcal{A}^{N,p}\left(\frac{f}{2\pi}\right)~,
\end{equation}
where above the appearance of the $U(1)$ field strength $f$ indicates that the bulk is coupled to the TQFT by gauging its one-form symmetry using the magnetic one-form symmetry generated by $f$.  This mimics the familiar coupling of the electromagnetic gauge field in the fractional quantum hall effect. (See e.g.\ \cite{Kapustin:2014gua} for further discussion of coupling a TQFT to a QFT.)

The existence of the symmetry defects $\mathcal{D}_{k}$ may be viewed as a non-perturbative intrinsic definition of the abelian ABJ anomalous chiral symmetry valid at the operator level in the associated quantum field theory. As we discuss in detail in section \ref{selec} these symmetry defects impose selection rules on correlation functions.  For local operator correlation functions in flat space, these are the naive selection rules of the chiral symmetry.  However, for more general spacetime manifolds, or correlation functions involving extended operators, the selection rules are modified, but nevertheless encoded by the algebraic properties of $\mathcal{D}_{k}.$

Focusing on specific examples, we exhibit the defects $\mathcal{D}_{k}$ in massless quantum electrodynamics, and axion-electrodynamics.  In particular, the former implies the existence of novel non-invertible symmetries in the vanishing fermion mass limit of the standard model. We also exhibit examples of the symmetry defects $\mathcal{D}_{k}$ in non-abelian gauge theory, reviewing the construction of \cite{Kaidi:2021xfk}, as well as axion-Yang-Mills.

Like all symmetries, the non-invertible chiral symmetry defects we construct are scale invariant and preserved under renormalization group flow. We thus anticipate that these defects will find broad application in investigating the dynamics of quantum field theories.

\subsection{Symmetry Breaking and Hierarchies}

Exact global symmetries in quantum field theory provide exact selection rules on correlation functions.  However, symmetry can also be useful when it is broken by small effects.  In that context, the associated degeneracies and selection rules are weakly violated but are still protected from large corrections in effective field theory.  This is known as \emph{technical naturalness} \cite{tHooft:1979rat}.

In section \ref{monosec} below, we explore the paradigm of weakly broken non-invertible symmetry.  We focus on the discrete chiral symmetry defects $\mathcal{D}_{k}$ and exhibit an interplay between the semiclassical physics of magnetic monopoles and technically natural non-perturbative effects.

The key idea is to utilize the fact that in effective field theory, the magnetic one-form symmetries that featured in the construction of the defects $\mathcal{D}_{k}$ cannot be violated by any local operator deformation.  In the case of the magnetic one-form symmetry of abelian gauge theory this is transparent from the fact that the field strength $f$ is closed by the Bianchi identity.  This remains true in the presence of any perturbative interactions or charged matter.  Similarly, the discrete magnetic one-form symmetry of non-abelian gauge theory cannot be broken by operator deformations or the addition of charged matter.  

From a semiclassical point of view, this means that the breaking of these one-form symmetries can only proceed through non-perturbative physics or relatedly a change in the topology of field space.  In the case at hand, this is achieved by considering models where there are dynamical magnetic monopoles.  These monopoles can screen the 't Hooft line and so violate the one-form symmetry non-perturbatively.

There is a close connection between the physics of monopoles described above and the properties of instantons.  As we emphasize in section \ref{secABJ}, non-invertible chiral symmetries arise semi-classically when there are instantons which cannot be realized in $\mathbb{R}^{4}$ but can be realized in more complicated spacetime topologies or in the presence of extended operators.  From this point of view, the change in field space topology eluded to above is precisely what is needed to allow the instanton to be realized in $\mathbb{R}^{4}.$ 

Once the one-form symmetry is violated, the discrete chiral symmetry defects $\mathcal{D}_{k}$ are no longer topological and hence their selection rules are also broken.  This can be seen for instance from the fusion rules discussed below and derived in \cite{Choi:2021kmx, Kaidi:2021xfk, Choi:2022zal}.  
However, since the violation of the one-form symmetry is non-perturbative, so too is the violation of the $\mathcal{D}_{k}$ selection rules.  Thus, weakly broken one-form symmetry yields exponential small violations in the selection rules of the non-invertible symmetry defects $\mathcal{D}_{k}.$ Moreover,  these small corrections are technically natural since they are the leading terms violating $\mathcal{D}_{k}.$

In this way we construct a variety of technically natural models with exponential hierarchies in their effective field theory description.  This includes axion models (abelian or non-abelian) with exponentially suppressed potentials or energy splittings.  In this context we make contact with the recent work of \cite{Fan:2021ntg} which examined the interplay between axion potentials and magnetic monopoles.  We also discuss gauge theories with matter (abelian or non-abelian) with exponentially suppressed fermion masses.

We speculate that this new symmetry-based mechanism for generating exponential hierarchies may find applications in model building or in refining our understanding of naturalness in effective field theory.

\emph{Note Added:} While we were concluding this work, we were notified of \cite{Choi:2022jqy}, which independently constructs and studies the non-invertible chiral symmetry defects in massless QED.

\section{Non-Invertible Chiral Symmetry from Abelian ABJ Anomalies}\label{secABJ}

In this section we construct non-invertible symmetry defects in models with abelian ABJ-like anomalies.  Specifically we find a non-invertible defect for each chiral symmetry rotation by a rational angle.  As we discuss below, these non-invertible symmetry defects also make precise the fate of the chiral symmetry selection rules on correlation functions.

Throughout these examples, a key role is played by the bulk one-form symmetry whose current couples with the internal one-form symmetry of the topological degrees of freedom on the defect.  In the case of abelian gauge theory this is the continuous magnetic one-form symmetry  $U(1)^{(1)}$ with closed two-form current $\frac{f}{2\pi}.$ While, in the non-abelian gauge theory examples below it is a suitable discrete analog of this magnetic symmetry.

\subsection{Massless Electrodynamics}\label{qedsec}

Let us begin with abelian gauge theory with massless charged fermionic matter (i.e.\ massless QED) and gauge group $U(1)$. For simplicity we assume that we have $N_{f}$ species of massless electrons of unit electric charge, though our discussion readily generalizes to models with more complicated matter content; the essential feature is only the ABJ anomaly.

At the classical level this theory admits a chiral symmetry $U(1)_{\text{chiral}}$.  Denoting the positively (negatively) charged Weyl fermions as $\chi_{\alpha}^{\pm, a}$ with $a$ a flavor index, the chiral symmetry acts as:
\begin{equation}\label{thetadef}
\chi_{\alpha, +}^{a}\rightarrow e^{i\theta}\chi_{\alpha, +}^{a}~, \hspace{.2in}\chi_{\alpha, -}^{a}\rightarrow \chi_{\alpha, -}^{a}~.
\end{equation}
When quantum effects are considered, it is well known that this symmetry is violated by the Adler-Bell-Jeckiv anomaly \cite{Adler:1969gk, Bell:1969ts}.  At a technical level, we may view this as arising from an anomaly polynomial related to the anomalous variation in the action by the descent procedure:\footnote{The five-dimensional inflow action $\mathcal{A}_{5}$ obeys $d\mathcal{A}_{5}=2\pi \mathcal{I}_{6}$.} 
\begin{equation}\label{inflow}
\mathcal{I}_{6}=\frac{N_{f}}{2(2\pi)^3}f\wedge f\wedge F~,
\end{equation}
where above, $f$ is the field strength of the dynamical $U(1)$ gauge field, and $F$ is the background field strength of the putative chiral symmetry.  This anomaly polynomial breaks the classical $U(1)_{\text{chiral}}$ symmetry down to a $\mathbb{Z}_{N_{f}}$ discrete chiral symmetries which is free of $ABJ$ anomalies.  In a modern language, this $\mathbb{Z}_{N_{f}}$ symmetry is generated by codimension one topological operators with standard group-like (invertible) fusion rules.

What more can be said about this well-known story?  A clue can be seen via a familiar trick in current algebra.  Denoting by $J_{\mu}$ the anomalous current and $*J$ the hodge dual, the anomaly \eqref{inflow} implies:
\begin{equation}\label{currdiff}
d* J  = \frac{N_{f}}{8\pi^{2}} f\wedge f~.
\end{equation}
The right-hand side is globally well-defined closed four-form. However if the spacetime manifold is sufficiently simple, in particular if it does not have any closed two cycles, and there are no line operators inserted, then the form is also exact and is the exterior derivative of the Chern-Simons three-form $\propto a\wedge da$.  Thus, in such a simple configuration, one is tempted to redefine the current as
\begin{equation}\label{curr2}
*J\stackrel{?}{\rightarrow} *J-\left(\frac{N_{f}}{8\pi^{2}}\right)a\wedge da~.
\end{equation}
This modified current is not gauge invariant, but in these simple field configurations certain rational multiples (with denominator $N_{f}$) of the total exponentiated charge integrals are well-defined and conserved.  Thus, one is tempted to declare that the chiral symmetry remains, despite the anomaly \eqref{currdiff}.  For instance, this manipulation implies that the selection rules of the broken chiral symmetry are valid when considering correlation functions of local operators in a topologically trivial spacetime.  

With a view towards later generalizations, we can also rephrase this argument in terms of instantons.  The equation \eqref{currdiff} implies that the chiral charge is violated by abelian instantons, but on a topologically trivial spacetime such as $\mathbb{R}^{4}$ (or more precisely its IR regulated version, $S^{4}$), with only local operator insertions, no such instantons exist.

The preceding discussion begs the question as to what has become of the chiral symmetry in massless QED (or other similar models with an abelian ABJ anomaly).  As we will now show, the correct statement which holds on any spacetime manifold in any configuration of operators is that there is a non-invertible chiral symmetry in the theory.  Thus, the true effect of the ABJ anomaly is not to destroy the chiral symmetry, but to change its nature.

It is helpful to consider again the anomaly polynomial \eqref{inflow}.  Let us focus on a discrete chiral rotation by a root of unity so that the angle in \eqref{thetadef} is $\theta=2\pi/k$ for some integer $k$.  This is implemented by a codimension one domain wall operator, $\mathcal{C}_{k}$, located say at $x=0$, where the chiral background gauge field $A$ (with field strength $F$) is taken to be: 
\begin{equation}
A=\theta \delta(x)dx~.
\end{equation}
Applying inflow, we deduce that across this domain wall, the bulk action differs by an effective theta angle:
\begin{equation}\label{inflowwall}
S\rightarrow S+\frac{2\pi iN_{f}}{k}\int_{x>0}\frac{f\wedge f}{8\pi^{2}}~.
\end{equation}
This equation implies that in a theory where the dynamical gauge field is a frozen background, the $\mathbb{Z}_{k}$ chiral domain wall $\mathcal{C}_{k}$ has a worldvolume 't Hooft anomaly characterized by the above term by inflow.  Once the $U(1)$ is made dynamical, we then see from \eqref{inflowwall} that the bulk action jumps across $\mathcal{C}_{k}$ and hence this defect does not define a symmetry of our theory.

The key idea, following \cite{Kaidi:2021xfk}, is now to modify the $\mathbb{Z}_{k}$ chiral domain wall $\mathcal{C}_{k}$ by stacking it with a suitable 3d TQFT which also couples to the bulk gauge field $a$ and cancels the apparent anomaly in \eqref{inflowwall}.  After this construction an avatar of the discrete chiral domain wall will remain as a topological defect (symmetry) in the theory, but at the cost of becoming non-invertible. 

In more detail, we consider a TQFT with a one-form symmetry $\mathbb{Z}_{N}^{(1)}$ with associated background field $B^{(2)}.$  This means that among the lines characterizing this theory are abelian anyons whose fusion algebra forms the group $\mathbb{Z}_{N}$.  These lines have a spin, with $s$-times the generating line having spin $h(s)$ given by
\begin{equation}\label{spinp}
h(s)=\frac{ps^{2}}{2N}~~ \mod 1~,
\end{equation}
where $p$ is coprime to $N$. Such a topological field theory carries an anomaly defined by inflow as:
\begin{equation}
S_{\text{inflow}}=-\frac{2\pi i p}{2N}\int \mathcal{P}(B^{(2)})~,
\end{equation}
where $\mathcal{P}$ is the Pontryagin square operation.\footnote{The Pontryagin square is a cohomology operation 
\begin{equation}
\mathcal{P}:H^2(M,\mathbb{Z}_N)\to H^4(M,\mathbb{Z}_{\gcd(N,2)N})~.
\end{equation}
In the case, where $B^{(2)}\in H^2(M,\mathbb{Z}_N)$ admits an integral uplift $\tilde{B}^{(2)}\in H^2(M,\mathbb{Z})$ (the main case of interest in the present paper), we have $\mathcal{P}(B^{(2)})=\tilde B^{(2)}\wedge \tilde B^{(2)}~\mod 2N.$ } As discussed in detail in \cite{Hsin:2018vcg}, any such TQFT can be written as product of a minimal abelian TQFT $\mathcal{A}^{N,p}$ together with a decoupled sector that plays no role in the following.  We denote by $\mathcal{A}^{N,p}\left(B^{(2)}\right)$ this minimal TQFT coupled to its one-form symmetry background $B^{(2)}$.

We now fix $N$ and $p$ such that $p/N=N_{f}/k$ and $\gcd(p,N)=1$. We define a modified chiral symmetry defect $\mathcal{D}_{k}$, by stacking $\mathcal{C}_{k}$ with $\mathcal{A}^{N,p}$ and coupling to the bulk by gauging the one-form symmetry of the TQFT through the dynamical gauge field $f=da$.  In equations this means:
\begin{equation}\label{Dkdef}
\mathcal{D}_{k}\equiv \mathcal{C}_{k}\times \mathcal{A}^{N,p}\left(\frac{f}{2\pi}\right)~.
\end{equation}
Thus, the bulk and TQFT are coupled through the fact that the one-form symmetry on the defect is identified with the magnetic one-form symmetry of the $U(1)$ gauge theory.

The coupling between the bulk and the defect defined by \eqref{Dkdef} takes a more familiar form in the special case $p=1$ where $\mathcal{A}^{N,p}$ is an abelian Chern-Simons theory $U(1)_{N}$ \cite{Barkeshli:2014cna, Hsin:2018vcg}. Denoting the defect dynamical gauge field by $c$ the bulk-defect coupling arises from a mixed Chern-Simons term:
\begin{equation}
S_{\text{defect}}=\frac{iN}{4\pi} \int c \wedge dc+\frac{i}{2\pi}\int c \wedge f~,
\end{equation}
which mimics the familiar coupling of the electromagnetic gauge field in the fractional quantum hall effect.  This point of view also clarifies the meaning of the gauge non-invariant current appearing in \eqref{curr2}: the fractional Chern-Simons term appearing in the current may be understood as the effective response of the fractional hall state.

The defect $\mathcal{D}_{k}$ defines the remnant of the $\mathbb{Z}_{k}$ chiral symmetry in our model.  Repeating the argument for general $k$ or powers thereof we see that \emph{a chiral rotation by any rational angle can be promoted to a general non-invertible topological defect.}  In particular, all such operators should be viewed as a generalized symmetries and imply selection rules on correlation functions discussed below.  

The non-invertible nature of these defects is manifest simply from the fact that the partition function of $\mathcal{A}^{N,p}$ in general does not have unit modulus (and may even vanish).  For instance, wrapping $\mathcal{D}_{k}$ on a three-sphere and using $|Z_{\mathcal{A}^{N,p}}(S^{3})|=1/\sqrt{N}$, we obtain the quantum dimension 
\begin{equation}\label{qdim}
\mathcal{D}_{k}\left(S^{3}\right)=\frac{1}{\sqrt{N}}~.
\end{equation}
Alternatively, one can also exhibit the non-invertibility of these defects through their fusion algebra.   In particular, following the derivation in \cite{Choi:2021kmx, Kaidi:2021xfk, Choi:2022zal}. one can deduce that the fusion of the defect $\mathcal{D}_{k}$ with its orientation reversal $\overline{\mathcal{D}}_{k}$ on a general oriented three-manifold $M$ is:
\begin{equation}\label{fusionrules}
\mathcal{D}_{k}(M)\times \overline{\mathcal{D}}_{k}(M)=\sum_{S\in H^{2}(M,\mathbb{Z}_{N})}\eta(S)e^{\frac{2\pi\mathrm{i}p}{N}Q(S)}~, 
\end{equation}
where $\eta=\exp\left(i\int_{S}\frac{f}{2\pi N}\right)$ is the generator of the $\mathbb{Z}_N$ subgroup of the magnetic one-form symmetry, the sum is over two-cycles $S \subset M$, and $Q(S)$ is the triple self-intersection number.  In particular, the right-hand side of \eqref{fusionrules} is an example of a condensation defect explored in detail in \cite{Gaiotto:2019xmp, Choi:2022zal} (see also \cite{Else:2017yqj,Johnson-Freyd:2020twl}).

In summary, the existence of the non-invertible topological defects $\mathcal{D}_{k}$ is an intrinsic statement about the operator content of the quantum field theory.  Thus, these defects give a non-perturbative definition of the ABJ anomaly and its physical consequences.  

\subsection{Axion Electrodynamics}

The non-invertible defects $\mathcal{D}_{k}$ defined above have analogs in any theory with an ABJ-like anomaly.  Let us exhibit them in a model of axion electrodynamics.  We focus on a minimal model, though our analysis applies to more general field content.  

Thus, consider a theory where, in addition to the dynamical $U(1)$ gauge field with field strength $f,$ there is a dynamical periodic scalar field $\theta$ (the axion) with:
\begin{equation}
\theta\sim \theta +2\pi \mu~,
\end{equation}
with $\mu$ the axion decay constant.  The action is:
\begin{equation}\label{axionem}
S=\frac{1}{2}\int d\theta \wedge *d\theta+\frac{1}{2e^{2}}\int f\wedge * f+\frac{iL}{8\pi^{2}\mu} \int \theta f \wedge f~,
\end{equation}
with $L\in \mathbb{Z}$ a possible discrete coupling constant.  Ignoring the coupling between the $\theta$ and the gauge field, the axion enjoys a continuous shift symmetry with current $J$ where $*J=i\mu*d\theta$.  However this current is broken by the coupling between $\theta$ and $f$:\footnote{The normalization of the current $J$ can be verified by noting that the associated charged operators are $\exp(in\theta/\mu)$ for integer $n$.}
\begin{equation}\label{axioncurrent}
d*J=\frac{L}{8\pi^{2}}f\wedge f~.
\end{equation}
This equation is directly analogous to the violation of the chiral symmetry in abelian gauge theory \eqref{currdiff} and breaks the axion shift symmetry to $\mathbb{Z}_{L}$.  However, as in our previous discussion we note that the violation of the axion shift symmetry is only through abelian instantons, and hence does not occur in sufficiently simple spacetime topologies with only local operator insertions.  We thus anticipate that \emph{the axion shift symmetry is not broken, but instead is transformed into a non-invertible topological operator.}

We may directly verify this intuition by constructing a suitable symmetry defect $\mathcal{D}_{k}$ associated to a discrete $\mathbb{Z}_{k}$ rotation of the axion $\theta\rightarrow \theta+\frac{2\pi \mu}{k}.$  Letting $\mathcal{C}_{k}$ denote the naive (broken) discrete $\mathbb{Z}_{k}$ shift symmetry defect, we see that across $\mathcal{C}_{k}$ the action shifts as in \eqref{inflowwall}, effectively generating a rational theta-angle $2\pi L/k$ for the dynamical $U(1)$ gauge field.  Again choosing $N$ and $p$ coprime with $p/N=L/k$ we define $\mathcal{D}_{k}$ exactly as in \eqref{Dkdef}:
\begin{equation}
\label{Dkdef2}
\mathcal{D}_{k}\equiv \mathcal{C}_{k}\times \mathcal{A}^{N,p}\left(\frac{f}{2\pi}\right)~.
\end{equation} 

Of course the close analogy between the axion shift symmetry and the chiral symmetry in abelian gauge theory is no accident as the two can be related by renormalization group flow. Indeed, in a typical model of axions constructed by the Pecci-Quinn mechanism, the UV consists for instance of electrodynamics coupled to a neutral complex scalar $\varphi$ and $N_{f}$ electrically charged fermions $\chi_{ \pm}^{a}$ \cite{Kim:1979if,Shifman:1979if}.  In addition to the kinetic terms the action contains Yukawa and potential interactions:  
\begin{equation}
\label{spq}
S\supset \int d^4x\big\{ \lambda \bar\varphi \chi_{+}^{a}\chi_{-}^{a}+\lambda \varphi  \bar{\chi}_{+}^{a}\bar{\chi}_{-}^{a}-V(\varphi)\big\}~.
\end{equation}
At the classical level, there is a chiral $U(1)_{PQ}$ symmetry where the different fields have charges 
\begin{center}
\begin{tabular}{c|c}
Field&$U(1)_{PQ}$ charge\\
\hline
$\chi_+^{a}$&$ +1$\\
$\chi_-^{a}$&$ 0$\\
$\varphi$&$+1$ \\
\end{tabular}
\end{center}
At the quantum level, this $U(1)_{PQ}$ has an ABJ anomaly and hence following the discussion above, gives rise to a non-invertible discrete chiral symmetry defect $\mathcal{D}_{k}$.  

We now flow to axion electrodynamics by condensing the scalar with $\langle|\varphi|^2\rangle =\mu^{2}$.  This freezes the radial mode of $\varphi$.  However, an axion $\theta$ remains as a pseudogoldstone mode where
\begin{equation}
 \varphi=\mu \exp(i\theta/\mu)~.
 \end{equation}  
The Yukawa couplings give a mass to the fermions, and decoupling them from the axion requires a chiral $U(1)_{PQ}$ rotation.   The anomaly then generates an axion coupling between $\theta$ and the $U(1)$ gauge fields.  Thus, the final action is that of axion-electrodynamics where the coupling $L$ in \eqref{axionem} is identified with $N_{f}$ the number of UV flavors.  

This simple flow illustrates the renormalization group invariance of the existence of the non-invertible symmetry defects. Indeed, the defect $\mathcal{D}_{k}$ constructed from the discrete chiral symmetry in the UV flows to the non-invertible axion shift symmetry defect in the IR.

In general, the defect $\mathcal{D}_{k}$ defined in \eqref{Dkdef2} makes rigorous the sense in which the axion shift symmetry is still present in axion electrodynamics.  As discussed in more detail below this means that at the level of local operator correlation functions on $\mathbb{R}^{4}$, the selection rules implied by the discrete axion shift symmetry hold.  One important application of this is to the axion potential.  Taking into account only the standard $\mathbb{Z}_{L}$ invertible symmetry, the axion model admits a possible potential of the form:
\begin{equation}
V(\theta)=\sum_{n \in \mathbb{N}}\alpha_{n}\cos\left(\frac{Ln\theta}{\mu}\right)~.
\end{equation}
However, this potential is forbidden by the non-invertible defect $\mathcal{D}_{k}$ which shifts $\theta$ by a fraction of its period.  In other words a vanishing potential in axion electrodynamics is stabilized by the exact non-invertible symmetry generated by the defects $\mathcal{D}_{k}$.

\subsection{Non-Abelian Gauge Theory with Matter}

A discrete part of the non-invertible symmetry in the previous sections can also arise in a non-abelian gauge theory.
This symmetry was discussed in \cite{Kaidi:2021xfk}, and here we recast the argument so that it aligns with the ABJ-anomalous symmetry presented above.  We focus on the simplest example of relevance to later models, though as in previous sections the essential feature is the anomaly.

Consider e.g.\ a QFT with nonabelian gauge group $PSU(N_c)\cong SU(N_{c})/\mathbb{Z}_{N_{c}},$ with non-abelian field strength $w$. We take the matter content to be $N_f$ Weyl fermions in the adjoint representation, $\lambda^a,$ where $a$ is a flavor index. There is a classical $U(1)_{\text{chiral}}$ symmetry acting on $\lambda^a$ as $\lambda^a \to e^{\mathrm{i}\theta}\lambda_a$.  In this case the anomaly polynomial is 
\begin{equation}\label{inflow_nonab}
\mathcal{I}_{6}=\frac{2N_{f}N_c}{2(2\pi)^3}\mathrm{Tr}(w\wedge w)\wedge F~,
\end{equation}
where $\mathrm{Tr}$ is the trace in the fundamental representation, and $F$ is again the background $U(1)_{\text{chiral}}$ field strength.

The resulting pattern of discrete chiral symmetry depends on the global form of the gauge group through the properties of the instanton number.  Specifically, (restricting to spin manifolds for simplicity), we have:
\begin{equation}\label{iquant}
\mathcal{I}=\frac{1}{8\pi^2}\int\mathrm{Tr}(w\wedge w)\in \begin{cases} \mathbb{Z}&SU(N_c)~,\\  \frac{1}{N_c}\mathbb{Z} &PSU(N_c)~. \end{cases}
\end{equation}
This means the invertible symmetry group is broken down to $\mathbb{Z}_{2N_cN_f}$ for the $SU(N_c)$ theory, and to $\mathbb{Z}_{2N_f}$ for the $PSU(N_c)$ theory.

The $PSU(N_c)$ configuration with fractional $\mathcal{I}$ is called a fractional instanton. Such a configuration is possible because a $PSU(N_c)$ gauge connection $a,$ has more freedom than a $SU(N_c)$ gauge connection.  Specifically, this is encoded in the discrete magnetic flux, defined cohomologically by the second Stiefel Whitney class of the gauge bundle $w_2(a) \in H^2(M,\mathbb{Z}_{N_c})$.  The fractional part of the instanton number $\mathcal{I}$ is controlled by $w_2(a)$ as:\footnote{This is an integer multiple of $\frac{1}{N_c}$ because, assuming spacetime is a spin manifold, $\mathcal{P}(w_2(a))$ is even when $N_c$ is even.}
\begin{equation}
    \mathcal{I} = \frac{N_c-1}{2N_c}\mathcal{P}(w_2(a)) \mod 1~.
\end{equation}
This fractional instanton breaks the anomaly-free symmetry $\mathbb{Z}_{2N_cN_f}$ of the $SU(N_c)$ theory further down to $\mathbb{Z}_{2N_f}$.

However, exactly analogous to our discussion of the ABJ anomaly for the chiral symmetry of massless QED, fractional instanton configurations do not exist when the space-time manifold has a trivial second cohomology group, $H^2(M,\mathbb{Z}_{N_c})$.  The main difference is that, in the abelian case, no instantons exist at all on  $\mathbb{R}^4$, while for $PSU(N_c)$ only the fractional instantons are absent on $\mathbb{R}^{4}$. Thus, we expect that the generator of $\mathbb{Z}_{2N_cN_f}$ survives as a topological defect, but becomes a non-invertible symmetry.

Proceeding as in the previous sections, we can construct the desired symmetry defect $\mathcal{D}_{k}$ with $k=2N_cN_f$ again by stacking the $\mathbb{Z}_{k}$ domain wall $\mathcal{C}_{k}$ with a minimal abelian TQFT as
\begin{equation}\label{nonabstack}
    \mathcal{D}_k = \mathcal{C}_k \times \mathcal{A}^{N_c,1}(w_2(a))~,
\end{equation}
where the TQFT couples to the Stiefel-Whitney class $w_2(a)$ instead of the field strength. This is the non-invertible defect found in \cite{Kaidi:2021xfk}.

\subsection{Axion Yang-Mills}\label{aym}

Just as the non-invertible chiral symmetry defects in massless QED immediately generalize to non-invertible symmetries of axion-electrodynamics, so too do the non-invertible symmetries described in the previous subsection generalize to non-invertible symmetries of axion-Yang-Mills.  Here we focus on examples with gauge group $PSU(N_{c})$ to parallel the above discussion, though it is straightforward to generalize to other gauge groups.  

Thus we consider a QFT, where in addition to the kinetic terms involving the axion, $\theta,$ and the gauge field strength $w$, the action includes the interaction
\begin{equation}\label{nonabact}
S \supset \frac{iL}{8\pi^{2}\mu}\int \theta \  \mathrm{Tr}(w\wedge w)~.
\end{equation}
Here we must carefully keep track of the quantization of the instanton density expressed in \eqref{iquant}.  In particular, fractional instantons of $PSU(N_{c})$ imply that the correct periodicity of the axion is:
\begin{equation}\label{fullperiod}
\theta\sim \theta +2N_{c}\pi \mu~.
\end{equation}

As discussed above the possibility of fractional instantons also leads to a discrete non-invertible symmetry defects.  Indeed, the model \eqref{nonabact} has a $\mathbb{Z}_{L}$ standard invertible symmetry which acts by shifting $\theta$ by $1/L$ times its full periodicity.  However, there is also a non-invertible topological defect $\mathcal{D}_{k}$ with $k=N_{c}L$ constructed as in the previous section by stacking a discrete axion shift symmetry $\mathcal{C}_{k}$ domain wall, with the minimal TQFT $\mathcal{A}^{N_{c},1}(w_2(a))$ as in \eqref{nonabstack}.  

The non-invertible defect $\mathcal{D}_{k}$ contains the invertible defects by fusion in the sense that $(\mathcal{D}_{k})^{N_{c}}$ is the generator of the invertible $\mathbb{Z}_{L}$ symmetry.  As in the axion-electrodynamics case, $\mathcal{D}_{k}$ constrains the axion potential by enforcing that it is invariant under shifts of by $1/(N_{c}L)$ of the full period: $\theta\rightarrow \theta+\frac{2\pi \mu}{L}.$  
In particular, the potential has $N_cL$ minima leading to $N_cL$ degenerate vacua.

\subsection{Selection Rules on Correlation Functions }\label{selec}

An ordinary invertible symmetry imposes selection rules; it ensures that some correlation functions vanish, or more generally relates one correlation function to another. It is thus natural to ask what constraints on correlation functions are implied by the non-invertible topological defects $\mathcal{D}_{k}$ defined above.

\begin{figure}[t]
    \centering
        \includegraphics[width=\linewidth]{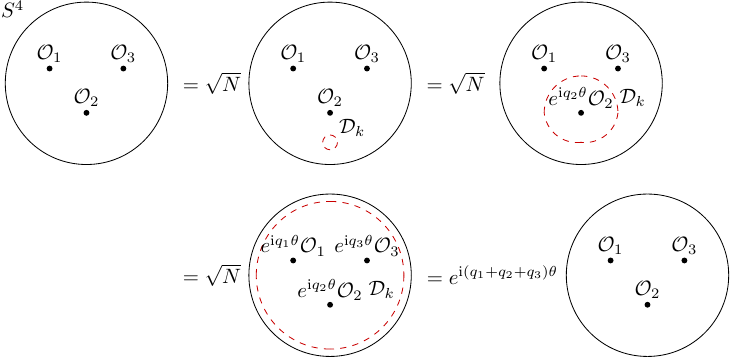}
    \caption{The action of the non-invertible symmetry $\mathcal{D}_k$ on the correlation function on $S^4$ with local operator insertions with charge $q_1,q_2$, and $q_3$. The rotation angle $\theta$ of the symmetry is $\frac{2\pi}{k}$. At the last step the defect is contracted at the other side of the sphere.}
    \label{fig:action}
\end{figure}

When we consider correlation functions of local operators on $S^4$ (regarded as an IR regulated $\mathbb{R}^4$), the selection rule is simple: we just ignore the fact that the defect is non-invertibile and treat it as if it were an ordinary (i.e.\ invertible) chiral symmetry (see also \cite{Harlow:2018tng} for related discussion).  To formally derive this result, we insert a small $S^3$-shaped symmetry defect $\mathcal{D}_k (S^3)$ around the north pole of the ambient $S^4$, then let the defect pass through the whole spacetime, contracting at the south pole (see Fig.~\ref{fig:action}). In this procedure the local operators acted on by the $\mathcal{C}_k$ part of $\mathcal{D}_k$, but do not interact with the TQFT $\mathcal{A}^{N,p}.$\footnote{When we insert the defect $\mathcal{D}_k (S^3)$ we have to divide the correlator by a factor of the quantum dimension \eqref{qdim}, however this factor cancels when we contract the defect at the antipodal point.}

The selection rules become more interesting when either the manifold is more complicated, or we involve line operators in the correlator.
Let us focus on the latter.  Consider for example a 't Hooft line operator $T$ of magnetic charge $m$ inserted along a curve $\gamma$ in spacetime.  In and an abelian gauge theory with symmetry $\mathcal{D}_{k}$ defined in \eqref{Dkdef}, this operator may be viewed as the worldline of an infinitely heavy Dirac monopole.

How does the non-invertible symmetry act on  $T$?  In this case the TQFT degrees of freedom of the defect $\mathcal{D}_k$ are essential.  Examining \eqref{inflowwall}, we see that the domain wall effectively changes the theta term as \eqref{inflowwall}.  This implies that the line $T$ acquires a fractional electric charge through the Witten effect.  Hence the action on $T$ is 
\begin{equation}
    T(\gamma) \to T(\gamma) W^{\frac{N_f}{k}}(\gamma,\Sigma)~,
\end{equation}
where $W^{\frac{N_f}{k}}(\gamma,\Sigma) = e^{\mathrm{i}\frac{N_f}{k}\int_\Sigma f}$ with $\partial \Sigma = \gamma$, is an open one-form symmetry generator. When $\frac{N_f}{k}$ is an integer, the operator $W^{\frac{N_f}{k}}(\gamma,\Sigma)$ is a genuine Wilson line operator with the integer charge, but in general for fractional $\frac{N_f}{k}$ it depends on the surface $\Sigma$.

This behavior of the non-invertible symmetry, transforming a genuine line operator to a line attached to a surface, was also found in \cite{Choi:2021kmx, Kaidi:2021xfk, Choi:2022zal}.  It is a direct analog of Kramers Wannier duality, which exchanges the local spin operator and the disorder operator in the two-dimensional Ising model.

\begin{figure}[t]
    \centering
        \includegraphics[width=\linewidth]{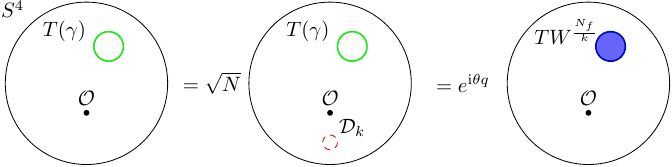}
    \caption{A non-invertible symmetry transforms a line operator $T$ on a line $\gamma$ into a line operator attached to a surface operator.}
    \label{fig:action_line}
\end{figure}

Although we have derived the properties of the defect action $\mathcal{D}_{k}$ in abelian QED, the result is model independent and intrinsically encoded in the interplay between the bulk one-form symmetry and the TQFT.  Thus, in general, the non-invertible defect $\mathcal{D}_k$ acts on a line charged under the one-form symmetry, and transforms it into a non-genuine line attached to an open one-form symmetry surface operator.    An example of this action on a correlator involving $T$ and local operators on $S^4$ is shown in Figure~\ref{fig:action_line}. Note that in the presence of $T$ insertions, non-zero correlation functions where the ABJ-anomalous charge is classically unbalanced are compatible with the $\mathcal{D}_{k}$ selection rules.

The non-invertibility of $\mathcal{D}_k$ can also be seen in its action on the Hilbert space of states on a non-trivial spatial manifold.
To be concrete, take the spatial manifold to be $M_3=S^2\times S^1$.
The Hilbert space decomposes into  sectors labelled by the one-form charge:
$\mathcal{H}_{S^2\times S^1} = \oplus_m \mathcal{H}_m$, where $m = \int_{S^2}\frac{f}{2\pi}$. 
To study the action of $\mathcal{D}_k$ on this Hilbert space, we need the partition function of the TQFT $\mathcal{A}^{k,1}$ on this spatial manifold.
As the TQFT is coupled with $f$ through its one-form symmetry, the magnetic flux $m$ induces an anyon line along the $S^1$ direction of $M_3$, which is the $m$-th power of the generating line. However, the $S^2\times S^1$ partition function of $\mathcal{A}^{k,1}$ with an anyon insertion along $S^1$ vanishes unless the line is trivial. Therefore we deduce:
\begin{equation}
    \mathcal{A}^{k,1}[S^2\times S^1,m] = 
    \begin{cases}
        0 & m \neq 0 \mod k\\
        1 & m = 0 \mod k
    \end{cases}.
\end{equation}
From the definition \eqref{Dkdef}, we have
\begin{equation}
    \mathcal{D}_k[S^2\times S^1] = \mathcal{C}_k P_{m,k},
\end{equation}
where $P_{m,k}$ is the projection onto the sectors $\mathcal{H}_m$ with $m = 0 \mod k$.
The non-invertibility of $\mathcal{D}_k$ is now manifest from the fact that its action contains a projection operator.  Note also that preservation of this operator means that, on $\mathcal{H}_m$, the chiral central charge is preserved modulo $m$ if $m\neq0$, and it is exactly preserved on $\mathcal{H}_0$.

\section{Exponential Hierarchies from Magnetic Monopoles}\label{monosec}

The previous section constructed non-invertible analogs of chiral symmetry in gauge theory.  Here we consider models where these symmetries are emergent at long distances and show that this scenario naturally leads to exponential hierarchies in the violation of the chiral symmetry selection rules.  

As emphasized in section \ref{intro}, the unifying feature of these examples is that the breaking of a magnetic one-form symmetry proceeds semiclassically through the existence of magnetic monopoles which may screen the 't Hooft lines.  This breaking is communicated to the non-invertible chiral symmetry defects $\mathcal{D}_{k}$ and leads to non-perturbative, but technically natural, violations of the selection rules.

\subsection{Exponentially Suppressed Axion Potentials}

As a first pedagogical example, we consider a model of axion electrodynamics where the $U(1)$ gauge field arises from an ultraviolet non-abelian gauge group by Higgsing.  As is familiar, in this model one finds an exponentially small axion potential generated by instantons.  Following the analysis of \cite{Fan:2021ntg}, we recast this non-perturbative potential as the consequence of quantum loops of dynamical magnetic monopoles.  These monopoles break the magnetic one-form symmetry of the infrared theory.  In this way we recast the exponentially supressed axion potential as a technically natural violation of the non-invertible axion shift symmetry defects $\mathcal{D}_{k}$.

For concreteness, we focus on the simplest case of $SU(2)$ though the discussion readily generalizes.  At high energies, our effective description consists of a non-abelian gauge field with field strength $w$, the axion $\theta$, and a real $SU(2)$ triplet Higgs field $\Phi$.  In addition to the kinetic terms for $\theta,$ $w,$ and $\Phi,$ the action contains the interaction terms for the axion and gauge field, as well as the Higgsing potential
\begin{equation}\label{UVaxion}
S \supset \frac{iL_{UV}}{8\pi^{2}\mu}\int \theta \  \mathrm{Tr}(w\wedge w) -\frac{1}{g^{2}}\int d^{4}x ~V(\Phi)~.
\end{equation}
We assume the Higgsing potential is chosen such that the triplet condenses, $\langle |\Phi|^{2}\rangle = v^{2}$ leading at long distances to axion electrodynamics and a neutral decoupled scalar with $L_{IR}=2L_{UV}$.  Henceforth for simplicity, we set $L_{UV}\rightarrow1$.

Let us track the symmetries along this renormalization group flow.  In the IR we have the non-invertible axion shift symmetry defects $\mathcal{D}_{k}$.  However, in this model these symmetries emergent and explicitly broken by the couplings of the ultraviolet action \eqref{UVaxion}.  A simple way to understand this is again to look at the divergence of the semiclassical current $J$ that acts to shift the axion.  In the non-abelian model \eqref{UVaxion} we have:
\begin{equation}
d*J=\frac{1}{8\pi^{2}}\mathrm{Tr}(F\wedge F)~.
\end{equation}
Unlike the case of axion-electrodynamics, the right-hand side of this equation is not in general $d$-exact, even in $\mathbb{R}^{4}$, and hence no modified charge is available.  Said differently, there are non-trivial instantons in $S^{4}$ which explicitly violate the axion shift symmetry. 

One can also ask more directly why the construction of the defect $\mathcal{D}_{k}$ fails in the non-abelian gauge theory.  In this case as we cross the shift symmetry wall $\mathcal{C}_{k},$ the bulk differs by a rational non-abelian theta-angle $2\pi/k$.  However now we can no longer absorb this anomaly by dressing the defect by a suitable topological field theory.  Indeed, our coupling to $\mathcal{A}^{N,p}$ is through the one-form symmetry of the TQFT and the magnetic one-form symmetry of the bulk gauge theory.  But in the non-abelian theory this magnetic one-form symmetry is explicitly broken.

The preceding comments allow us to investigate the breaking of the defect $\mathcal{D}_{k}$ from the point of view of the IR abelian gauge theory.  In this theory, when the 't Hooft lines are viewed a rigid defect operators, the magnetic one-form symmetry is preserved. However, when there are dynamical magnetic monopoles, the lines can be screened and the one-form symmetry is broken.  This implies that we can understand the leading symmetry breaking effects for $\mathcal{D}_{k}$ from the semiclassical physics of magnetic monopoles.
In other words the dynamical monopole/dyon world lines generated in the vacuum appear as insertions of Wilson-'t Hooft line operators, where the selection rules discussed in section \ref{selec} do not ensure the vanishing of correlation functions.

Let us focus on the calculation of the effective axion potential which arises from summing over loops of monopoles in the abelian gauge theory. Our treatment follows that of \cite{Fan:2021ntg}.  Assuming the interactions between multiple monopoles are small, the effective potential $V(\theta)$ can be usefully computed in the worldline formalism where we sum over closed trajectories of particles with a unit magnetic charge. This can be viewed as a dilute monopole gas approximation.  Working at a constant value $\theta$ of the axion we then have
\begin{equation}
\int d^{4}x~V(\theta)=Z(\theta)~,
\end{equation}
where $Z(\theta)$ is the single particle partition function.  In turn, this partition function may be expressed as an integral over worldline proper time $\tau$.  This results in a standard expression:
\begin{equation}\label{veff1}
V(\theta)\sim \int_{0}^{\infty}d\tau ~\tau^{\alpha}\exp\left(-\frac{m^{2}(\theta)}{2}\tau\right)~,
\end{equation}
where $m(\theta)$ is the mass of the magnetically charged particle and we have neglected order one coefficients and a power law term in the integrand.  Thus, we see that an effective potential will arise for the axion $\theta$ when the particles summed in the worldline trajectory have masses that depend on $\theta$.  This, may be viewed as an analog of the Coleman-Weinberg potential \cite{Coleman:1973jx} generalized to include loops of solitons. (See also \cite{Kawasaki:2015lpf, Nomura:2015xil, Kawasaki:2017xwt} for related discussion.)

To evaluate \eqref{veff1} we use the fact that in addition to a fundamental monopole, there is also a tower of dyons whose electric charges $q$ depend on the axion through the Witten effect \cite{Witten:1979ey}:
\begin{equation}\label{witteneff}
q=\left(\ell-\frac{\theta}{2\pi \mu}\right)~,
\end{equation}
where $\ell \in \mathbb{Z}$ labels the dyon.  The energy spectrum of these dyons and their interactions can usefully be understood by viewing them as the result of quantizing a worldline periodic scalar where the axion $\theta$ enters as the worldline theta-angle \cite{Fischler:1983sc}. This results in a familiar energy spectrum for these modes 
\begin{equation}\label{roter}
E_{\ell}\sim \left(\ell-\frac{\theta}{2\pi \mu}\right)^{2}~,
\end{equation}
Equations \eqref{witteneff} and \eqref{roter} suggest a BPS ansatz where the $\theta$ dependence of the mass is taken to be 
\begin{equation}
m^{2}_{\ell}(\theta)\approx m_{\text{mon}}^{2}+m_{\text{vec}}^{2}\left(\ell-\frac{\theta}{2\pi\mu}\right)^{2}~, 
\end{equation}
with the mass of the monopole and W-boson respectively given by:
\begin{equation}
m_{\text{mon}}=4\pi v/g~, \hspace{.2in}m_{\text{vec}}=g v~.
\end{equation}
The effective potential \eqref{veff1} is then a sum over dyon species in addition to the integral over proper time.
\begin{eqnarray}
V(\theta)\hspace{-.05in}& \sim &\hspace{-.05in} \sum_{\ell \in \mathbb{Z}}\int_{0}^{\infty}\frac{d\tau}{ \tau^{\alpha}}\exp\left\{-\frac{\tau}{2}(m_{\text{mon}}^{2}+m_{\text{vec}}^{2}\left(\ell-\frac{\theta}{2\pi\mu}\right)^{2})\right\} \nonumber\\
\hspace{-.05in}& \sim &\hspace{-.05in}\sum_{n \in \mathbb{Z}} \int_{0}^{\infty}\frac{d\tau}{ \tau^{\beta}}\exp\left\{-\frac{m_{\text{mon}}^{2}\tau}{2}-\frac{2\pi^{2}n^{2}}{m_{\text{vec}}^{2}\tau}+\frac{in\theta}{\mu}\right\}~,
\end{eqnarray}
where above we have used Poisson ressumation.  Finally, evaluating the proper time integral in a saddle point approximation yields a critical proper time 
\begin{equation}
\tau_{*}=\frac{2\pi n}{m_{\text{mon}}m_{\text{vec}}}~,
\end{equation}
and a corresponding potential:
\begin{eqnarray}\label{vefffinal}
V(\theta) & \approx & \sum_{n \in \mathbb{N}} c_{n}\exp\left(-\frac{2\pi n m_{\text{mon}}}{m_{\text{vec}}}\right)\cos\left(\frac{n\theta}{\mu}\right)~, \nonumber \\
& \approx &  \sum_{n \in \mathbb{N}} c_{n}\exp\left(-\frac{8\pi^{2}n}{g^{2}}\right)\cos\left(\frac{n\theta}{\mu}\right)~.
\end{eqnarray}
Thus loops of monopoles generate an effective potential for the axion which is exponentially suppressed. 

The result \eqref{vefffinal} can also be understood from the ultraviolet point of view as the familiar axion potential generated by non-abelian instantons.  Indeed the weight of each term in the sum \eqref{vefffinal} is precisely the non-abelian instanton action.  What the presentation in terms of monopoles has accomplished is it has made manifest the link between one-form symmetry breaking via dynamical monopoles and the exponentially suppressed axion potential.  In particular, since this potential is the leading operator deformation violating the axion shift symmetry defects $\mathcal{D}_{k}$, it is technically natural.

\subsection{Exponentially Small Axion Energy Splittings}\label{energy}

The preceding example revisited a familiar small axion potential from the point of view of non-invertible symmetry, and presented a link to monopole physics and one-form symmetry breaking.   We can now use the same technique to generate more novel technically natural models with exponential hierarchies.

As an example we consider an axion-Yang-Mills  with infrared gauge group $SO(3)$.  In addition to the kinetic terms involving the axion, $\theta,$ and the gauge field strength $w$, the action includes the interaction
\begin{equation}\label{so3act}
S \supset \frac{i}{8\pi^{2}\mu_{IR}}\int \theta \  \mathrm{Tr}(w\wedge w)~.
\end{equation}
Noting that $SO(3)\cong PSU(2)$, the discussion of section \ref{aym} applies.  Specifically, the full periodicity of the axion is 
\begin{equation}\label{periodaxx}
\theta\sim \theta +4\pi \mu_{IR}~, 
\end{equation}
and there is a non-invertible symmetry defect  $\mathcal{D}_{2}$ which acts to shift the axion by half of its full period $\theta\rightarrow \theta+2\pi \mu_{IR}$.  In this case, the topological degrees of freedom on the defect are the semion Chern-Simons theory, $U(1)_{2}$.

The exact axion shift symmetry defect $\mathcal{D}_{2}$ implies exact selection rules on the axion potential.  Specifically, a general potential compatible with the full periodicity \eqref{periodaxx} takes the form:
\begin{equation}\label{irgen}
V(\theta)=\sum_{n\in \mathbb{N}}\alpha_{n} \cos\left(\frac{n\theta}{2\mu_{IR}}\right)~,
\end{equation}
where the coefficients $\alpha_{n}$ arise for instance from instantons which need not be suppressed due to strong non-abelian dynamics.  However, the symmetry defect $\mathcal{D}_{2}$ implies the selection rule:
\begin{equation}\label{selcoeff}
\alpha_{n}=0~, ~~~\text{if}~n~ \text{odd}~.
\end{equation}
This in turn leads to two ground states where $\langle \theta\rangle=0$ or $\langle \theta\rangle=2\pi \mu_{IR},$ whose exact degeneracy is protected by the topological defect $\mathcal{D}_{2}$. 

We now embed this model as the long-distance limit of a flow from a larger gauge group which breaks the one-form symmetry in the ultraviolet.  Specifically, let the parent gauge group be $SU(3)$ with field strength $W$ and suppose there is in addition a complex Higgs field $\Phi$ transforming in the $\mathbf{6}$ of $SU(3)$ (the symmetric product of two fundamental $\mathbf{3}$'s).  In addition to the kinetic terms, the action includes the interactions:
\begin{equation}\label{UVaxion2}
S \supset \frac{i}{8\pi^{2}\mu_{UV}}\int \theta \  \mathrm{Tr}(W\wedge W) -\frac{1}{g^{2}}\int d^{4}x ~V(\Phi)~.
\end{equation}
Here $\mu_{UV}$ is the ultraviolet axion decay constant, in terms of which the axion has standard periodicity $\theta \sim \theta+2\pi \mu_{UV}$.

We now assume that the potential is chosen to condense the Higgs field so that (up to gauge transformations) the expectation value of the scalar $\langle\Phi\rangle =v\delta_{ij}$ (the $3\times 3$ identity matrix.). This expectation value Higgses the gauge group $SU(3)$ down to the subgroup that preserves $\delta_{ij}$, which is $SO(3)$.  At long-distances we thus obtain $SO(3)$ axion Yang-Mills, together with an adjoint scalar field.  

We must also keep track of the properties of the instanton density under this pattern of Higgsing.  Crucially, the index of embedding of $SO(3)$ inside $SU(3)$ is four \cite{Csaki:1998vv}.  This means that a single instanton of $SO(3)$ has instanton number four when embedded inside $SU(3)$.   In particular, this implies that the relation between the axion decay constants is:
\begin{equation}\label{murel}
\mu_{UV}=4\mu_{IR}~.
\end{equation}

Let us next turn to the axion potential generated by the flow.  Since the one-form symmetry of the infrared theory is broken by dynamical monopoles, we anticipate a non-perturbative violation of the $\mathcal{D}_{2}$ selection rules.  Indeed, assuming that the Higgs scale $v$ is chosen so that at the scale $v$ the $SU(3)$ theory is weakly coupled, $SU(3)$ instantons will generate a small change in the axion potential $\delta V(\theta)$ which admits an expansion of the form:
\begin{equation}
\delta V(\theta)=\sum_{n\in \mathbb{N}} \beta_{n}v^{4}\exp\left(-\frac{8\pi^{2}n}{g^{2}(v)}\right)\cos\left(\frac{n\theta}{\mu_{UV}}\right)~.
\end{equation} 
In particular, using \eqref{murel}, and comparing to the general IR expansion \eqref{irgen}, we observe that all Fourier modes are now present in $\delta V(\theta)$.\footnote{
    There are also naively modes violating the periodicity \eqref{periodaxx}. Consistency of the above analysis therefore requires that odd numbered instantons of $SU(3)$ do not contribute to the axion potential. A related vanishing of the contribution to the superpotential from odd numbered $SU(3)$ instantons in the same Higgsing pattern in a supersymmetric context was observed in \cite{Csaki:1998vv}. Both phenomena likely follow from the fact that in these models the dyon is a fermion.
}

We conclude that the selection rule \eqref{selcoeff} is now violated non-perturbatively, and hence there is an exponentially suppressed energy splitting between the nearly degenerate states:
\begin{equation}
|\delta V(0)-\delta V(2\pi \mu_{IR})|\sim v^{4}\exp\left(-\frac{8\pi^{2}}{g^{2}(v)}\right)~.
\end{equation}
Since this is the leading effect violating the defect $\mathcal{D}_{2}$, this exponential potential splitting is technically natural.

\subsection{Exponentially Suppressed Mass Terms in QED}

The previous examples used the breaking of non-invertible symmetries to construct axion potentials with interesting features.  Here we apply the same mechanism to produce a small breaking of the non-invertible chiral symmetry defect $\mathcal{D}_{k}$ in massless QED.  This results in technically natural models with exponentially suppressed fermion masses.

As the simplest example, let the ultraviolet be $SU(2)$ gauge theory with two Weyl fermions $\chi_\sga$ and $\psi_\sga$ in the fundamental representation ($\mathbf{2}$) where $\sga = 1,2$ is a doublet gauge index.\footnote{Note that the number of Weyl doublets must be even for consistency with the Witten anomaly \cite{Witten:1982fp}.}  In addition we also include a real adjoint scalar Higgs field $\Phi_{ij}$.  At short distances, $SU(2)$ instantons completely break the chiral symmetry rotating the fermions $\chi$ and $\psi$, except for the fermion parity $(-1)^F$ which is contained in the $SU(2)$ gauge group.  Anticipating the discussion to follow, we also note that this theory does not have any one-form symmetry.

We now use a potential $V(\Phi)$ to condense the scalar, $\langle\Phi_{\sga\sgaa}\rangle = v\delta_{\sga\sgaa}$. This Higgses the gauge group down to $SO(2)\cong U(1)$.  The long-distance theory has two fermions $\chi_+$ with charge $+1$ and two fermions $\chi_-$ with charge $-1$. 

If the bare mass term for the fermions vanishes in the ultraviolet, then at long distances this model will have emergent non-invertible chiral symmetries acting on the fermions of the type discussed in section \ref{qedsec}.  Crucially, these chiral symmetries forbid a fermion mass term.  However, since the one-form symmetry is broken at short distances, we expect that this non-invertible symmetry is broken non-perturbatively by monopole bubbles, generating an exponentially suppressed mass term.  This expectation can be directly verified using $SU(2)$ instanton calculus. The 't Hooft vertex coming from the sector of instanton number one generates a bilinear $\chi\psi$ term.  At long distances this  gives a mass correction suppressed by the instanton factor $\sim e^{-8\pi^{2}/g^{2}(v)}$.

Thus, this simple example generates a non-perturbative violation of the non-invertible chiral symmetry selection rules.  However it is unsatisfactory, and not technically natural, since there is no mechanism prohibiting a bare fermion mass term in the ultraviolet.

To remedy this, and construct a technically natural model with exponentially light fermion masses, we consider a slight generalization of the model above.  We now take the ultraviolet to be $SU(2)$ gauge theory with $2N_f$ Weyl fermions and also make the adjoint scalar $\Phi_{ij}$ complex.  We denote the first of the fermions doublets as $\chi$, and rest of the doublets as $\psi^\bfl$, with flavor index $\bfl=1,\cdots, 2N_f-1$.  We aim to flow at long distances to QED with one light electron $\chi_+$ and positron $\chi_-$.

To achieve this in a technically natural way, we impose on the ultraviolet model the following (invertible, ABJ-anomaly-free) symmetry
\begin{equation}
    \chi \to \zeta \chi~, \quad
    \psi^\bfl \to \tilde{\zeta}\zeta \psi^\bfl~, \quad
    \Phi \to (\tilde{\zeta}\zeta)^{-2}\Phi~,
\end{equation}
where $\zeta^{2N_f} = 1$ and $\tilde\zeta^{2N_f-1}=1$ are roots of unity. This transformation generates a $\mathbb{Z}_{2N_f(2N_f-1)}$ invertible symmetry.
This symmetry prohibits the bare mass for the fermions and the Yukawa coupling between $\phi$ and $\chi$.  However, it allows a Yukawa interaction of the form:
\begin{equation}\label{yukwa}
y_{\bfl\bfll}\phi_{\sga\sgaa}\psi^{\bfl,\sga}\psi^{\bfll,\sgaa}~,
\end{equation}
where $y_{\bfl\bfll}$ is a symmetric matrix of couplings.

We again Higgs to $U(1)$ by giving an expectation value $\langle\Phi_{\sga\sgaa}\rangle = v \delta_{\sga\sgaa}$.  The Yukawa term \eqref{yukwa} implies that the $2N_f-1$ $SU(2)$ fundamental fermions $\psi^\bfl$ get a mass $vy_{\bfl\bfll}$ and only one electron-positron pair $\chi = (\chi_+,\chi_-)$ remains light. Again, the mass term for $\chi$ is protected at long distances only by the non-invertible chiral symmetry which is broken by dynamical monopoles.  

Following our previous examples, we thus expect that $SU(2)$ instantons gives an exponentially suppressed mass to $\chi$.  To see this directly consider the 't Hooft vertex of the $SU(2)$ theory
\begin{equation}
    \frac{1}{v^{3N_f-4}}\exp\left(-\frac{8\pi^{2}}{g^{2}(v)}\right)\left(\chi\prod_\bfl \psi^\bfl\right)~.
\end{equation}
Contracting two of these 't Hooft vertices with the Yukawa interactions, we obtain an effective two instanton interaction
\begin{equation}\label{2inst}
    \frac{y^{2N_f-1}}{v^{2N_f-2}}\exp\left(-\frac{16\pi^{2}}{g^{2}(v)}\right)\chi^2(\Phi^\dagger)^{2N_f-1}~,
\end{equation}
which is neutral under the $\mathbb{Z}_{2N_f(N_f-1)}$ symmetry.  Here, the expression \eqref{2inst} is to be understood as a sum of all possible gauge contractions, and $y$ represents a typical value of elements in the symmetric Yukawa matrix $y_{\bfl\bfll}$ defined in \eqref{yukwa}.  Note also that when $N_f \ge 2$, this interaction term is irrelevant so that its bare value is naturally suppressed by a further UV (e.g.\ Planck) scale. 

We now determine the effect of the interaction \eqref{2inst} at long distances by  evaluating $\Phi$ at its expectation value $v \delta_{\sga\sgaa}.$  This produces an effective mass term generated by two $SU(2)$ instantons:
\begin{equation}
  vy^{2N_f-1}\exp\left(-\frac{16\pi^{2}}{g^{2}(v)}\right)\chi^2~.
\end{equation}
An alternative way to see this two-instanton contribution is to retain the massive fermions $\psi^\bfl$ in the IR theory.  Then the one-instanton 't Hooft vertex, when contracted with the Yukawa, gives a tiny fermion mixing term:
\begin{equation}
    v y^{2N_f-2}\exp\left(-\frac{8\pi^{2}}{g^{2}(v)}\right)\chi\psi^\bfl~,
\end{equation}
which in turn gives a tiny nonzero value to the lighter mass eigenvalue via the seesaw mechanism.

As with our models of axions, this is a technically natural suppression of the mass term, protected by the weakly broken non-invertible chiral symmetry.

\subsection{Exponentially Suppressed Mass Terms in QCD}

In the previous example the long-distance theory was abelian gauge theory with a non-invertible chiral symmetry which is defined for all rational angles.
Here we present a variant of the above model where the infrared theory is a non-abelian gauge theory whose non-invertible chiral symmetry protecting the fermion masses is broken by an ultraviolet instanton.

The short distance model is a $SU(N_c)$ gauge theory, which we Higgs down to $SO(N_c)$.  As matter content we take $N_f$ fundamental and anti-fundamental Weyl fermions, and two complex scalars $\Phi_1,\Phi_2$ in the representation with two symmetric fundamental gauge indices.
As before, we separate one of the fundamental fermions $\chi$ from the rest of fundamental fermions $\psi^\bfl$, $\bfl = 1, \cdots N_f-1$ and we indicate all of the anti-fundamental fermions as $\overline\psi^{\bfl'}$, $\bfl' = 1, \cdots N_f$.

On this ultraviolet model, we impose and invertible $\mathbb{Z}_{2N_f(N_f-1)}$ symmetry acting as:
\begin{equation}
    \begin{split}
    \chi &\to \zeta \chi~, \quad
    \psi^\bfl \to \tilde{\zeta}\zeta \psi^\bfl~,\quad
    \overline{\psi}^{\bfl'} \to (\tilde{\zeta}\zeta)\overline\psi^{\bfl'}~,\\
    \Phi_1 &\to (\tilde{\zeta}\zeta)^{2}\Phi_1~,\quad
    \Phi_2 \to (\tilde{\zeta}\zeta)^{-2}\Phi~,
    \end{split}
\end{equation}
with $\zeta^{2N_f} = 1$ and $\tilde{\zeta}^{2(N_f-1)} = 1$ are roots of unity generating the symmetry action.  This symmetry allows the Yukawa interactions $y_{\bfl\bfll}(\Phi_1)^\dagger\psi^\bfl\psi^\bfl$ and $y'_{\bfl'\bfll'}\Phi_2\overline{\psi}^{\bfl'}\overline{\psi}^{\bfll'}$.

By giving an expectation value to the Higgs fields $\langle\Phi^{\bar{i}\bar{j}}_{1,2}\rangle = v_{1,2}\delta^{\bar{i}\bar{j}}$, the theory is Higgsed down to a $SO(N_c)$ gauge theory.  There is a single light Weyl fermion $\chi$ in the vector representation of $SO(N_c),$ while the other fermions acquire a mass directly from the Yukawa interactions.\footnote{For simplicity below we do not distinguish the separate expectation values $v_{1}$ and $v_{2},$ indicating both of their scales as $v$.}

Note that when $N_c=3$, this pattern of Higgsing is the same as that investigated in section \ref{energy} and also discussed in \cite{Kaidi:2021xfk}. In particular, in infrared $SO(N_{c})$ gauge theory the discrete chiral symmetry $\chi \to \mathrm{i}\chi$ is non-invertible and forbids a mass term for $\chi$.  This holds for a general $N_c$.

In the parent $SU(N_{c})$ theory, dynamical monopoles break the one-form symmetry leading to a non-perturbative violation of the non-invertible chiral symmetry selection rules.  This results in a technically natural mass term for $\chi$ suppressed by an ultraviolet instanton factor.  

To see this directly, we again examine the $SU(N_{c})$ 't Hooft vertex:
\begin{equation}
    \frac{1}{v^{3N_f-4}}\exp\left(-\frac{8\pi^{2}}{g^{2}(v)}\right)\chi\prod_{\bfl=1}^{N_f-1} \psi^\bfl\prod_{\bfl'=1}^{N_f}\overline{\psi}^{\bfl'}~.
\end{equation}
As before, we can contract two such vertices and the Yukawa couplings to arrive the effective interaction:
\begin{equation}
    \frac{y^{2N_f-1}}{v^{2N_f-2}}\exp\left(-\frac{16\pi^{2}}{g^{2}(v)}\right)\chi^2\Phi_1^{N_f-1}(\Phi_2^\dagger)^{N_f}~.
\end{equation}
Evaluating at the expectation value of the Higgs fields, this turns into an exponentially suppressed mass term for the $SO(N_{c})$ vector fermion $\chi$
\begin{equation}
    v\, y^{2N_f-1}\exp\left(-\frac{16\pi^{2}}{g^{2}(v)}\right)\chi^2~.
\end{equation}
This tiny mass is protected by the non-invertible chiral symmetry.

\section*{Acknowledgments}
We thank L. Delacretaz, H. Fukuda, and J. Harvey for useful discussions. We also thank the authors of \cite{Choi:2022jqy} for sharing a draft of their work.  CC is supported by the US Department of Energy DE- SC0021432 and the Simons Collaboration on Global Categorical Symmetries. KO is supported in part by JSPS KAKENHI Grant-in-Aid, No.22K13969 and the Simons Collaboration on Global Categorical Symmetries.

\bibliography{Ref}

\begin{thebibliography}{39}
\expandafter\ifx\csname natexlab\endcsname\relax\def\natexlab#1{#1}\fi
\expandafter\ifx\csname bibnamefont\endcsname\relax
  \def\bibnamefont#1{#1}\fi
\expandafter\ifx\csname bibfnamefont\endcsname\relax
  \def\bibfnamefont#1{#1}\fi
\expandafter\ifx\csname citenamefont\endcsname\relax
  \def\citenamefont#1{#1}\fi
\expandafter\ifx\csname url\endcsname\relax
  \def\url#1{\texttt{#1}}\fi
\expandafter\ifx\csname urlprefix\endcsname\relax\def\urlprefix{URL }\fi
\providecommand{\bibinfo}[2]{#2}
\providecommand{\eprint}[2][]{\url{#2}}

\bibitem[{\citenamefont{Gaiotto et~al.}(2015)\citenamefont{Gaiotto, Kapustin,
  Seiberg, and Willett}}]{Gaiotto:2014kfa}
\bibinfo{author}{\bibfnamefont{D.}~\bibnamefont{Gaiotto}},
  \bibinfo{author}{\bibfnamefont{A.}~\bibnamefont{Kapustin}},
  \bibinfo{author}{\bibfnamefont{N.}~\bibnamefont{Seiberg}}, \bibnamefont{and}
  \bibinfo{author}{\bibfnamefont{B.}~\bibnamefont{Willett}},
  \bibinfo{journal}{JHEP} \textbf{\bibinfo{volume}{02}}, \bibinfo{pages}{172}
  (\bibinfo{year}{2015}), \eprint{1412.5148}.

\bibitem[{\citenamefont{Fuchs et~al.}(2002)\citenamefont{Fuchs, Runkel, and
  Schweigert}}]{Fuchs:2002cm}
\bibinfo{author}{\bibfnamefont{J.}~\bibnamefont{Fuchs}},
  \bibinfo{author}{\bibfnamefont{I.}~\bibnamefont{Runkel}}, \bibnamefont{and}
  \bibinfo{author}{\bibfnamefont{C.}~\bibnamefont{Schweigert}},
  \bibinfo{journal}{Nucl. Phys. B} \textbf{\bibinfo{volume}{646}},
  \bibinfo{pages}{353} (\bibinfo{year}{2002}), \eprint{hep-th/0204148}.

\bibitem[{\citenamefont{Frohlich et~al.}(2004)\citenamefont{Frohlich, Fuchs,
  Runkel, and Schweigert}}]{Frohlich:2004ef}
\bibinfo{author}{\bibfnamefont{J.}~\bibnamefont{Frohlich}},
  \bibinfo{author}{\bibfnamefont{J.}~\bibnamefont{Fuchs}},
  \bibinfo{author}{\bibfnamefont{I.}~\bibnamefont{Runkel}}, \bibnamefont{and}
  \bibinfo{author}{\bibfnamefont{C.}~\bibnamefont{Schweigert}},
  \bibinfo{journal}{Phys. Rev. Lett.} \textbf{\bibinfo{volume}{93}},
  \bibinfo{pages}{070601} (\bibinfo{year}{2004}), \eprint{cond-mat/0404051}.

\bibitem[{\citenamefont{Frohlich et~al.}(2007)\citenamefont{Frohlich, Fuchs,
  Runkel, and Schweigert}}]{Frohlich:2006ch}
\bibinfo{author}{\bibfnamefont{J.}~\bibnamefont{Frohlich}},
  \bibinfo{author}{\bibfnamefont{J.}~\bibnamefont{Fuchs}},
  \bibinfo{author}{\bibfnamefont{I.}~\bibnamefont{Runkel}}, \bibnamefont{and}
  \bibinfo{author}{\bibfnamefont{C.}~\bibnamefont{Schweigert}},
  \bibinfo{journal}{Nucl. Phys. B} \textbf{\bibinfo{volume}{763}},
  \bibinfo{pages}{354} (\bibinfo{year}{2007}), \eprint{hep-th/0607247}.

\bibitem[{\citenamefont{Bhardwaj and Tachikawa}(2018)}]{Bhardwaj:2017xup}
\bibinfo{author}{\bibfnamefont{L.}~\bibnamefont{Bhardwaj}} \bibnamefont{and}
  \bibinfo{author}{\bibfnamefont{Y.}~\bibnamefont{Tachikawa}},
  \bibinfo{journal}{JHEP} \textbf{\bibinfo{volume}{03}}, \bibinfo{pages}{189}
  (\bibinfo{year}{2018}), \eprint{1704.02330}.

\bibitem[{\citenamefont{Chang et~al.}(2019)\citenamefont{Chang, Lin, Shao,
  Wang, and Yin}}]{Chang:2018iay}
\bibinfo{author}{\bibfnamefont{C.-M.} \bibnamefont{Chang}},
  \bibinfo{author}{\bibfnamefont{Y.-H.} \bibnamefont{Lin}},
  \bibinfo{author}{\bibfnamefont{S.-H.} \bibnamefont{Shao}},
  \bibinfo{author}{\bibfnamefont{Y.}~\bibnamefont{Wang}}, \bibnamefont{and}
  \bibinfo{author}{\bibfnamefont{X.}~\bibnamefont{Yin}},
  \bibinfo{journal}{JHEP} \textbf{\bibinfo{volume}{01}}, \bibinfo{pages}{026}
  (\bibinfo{year}{2019}), \eprint{1802.04445}.

\bibitem[{\citenamefont{Thorngren and Wang}(2019)}]{Thorngren:2019iar}
\bibinfo{author}{\bibfnamefont{R.}~\bibnamefont{Thorngren}} \bibnamefont{and}
  \bibinfo{author}{\bibfnamefont{Y.}~\bibnamefont{Wang}}
  (\bibinfo{year}{2019}), \eprint{1912.02817}.

\bibitem[{\citenamefont{Thorngren and Wang}(2021)}]{Thorngren:2021yso}
\bibinfo{author}{\bibfnamefont{R.}~\bibnamefont{Thorngren}} \bibnamefont{and}
  \bibinfo{author}{\bibfnamefont{Y.}~\bibnamefont{Wang}}
  (\bibinfo{year}{2021}), \eprint{2106.12577}.

\bibitem[{\citenamefont{Komargodski et~al.}(2021)\citenamefont{Komargodski,
  Ohmori, Roumpedakis, and Seifnashri}}]{Komargodski:2020mxz}
\bibinfo{author}{\bibfnamefont{Z.}~\bibnamefont{Komargodski}},
  \bibinfo{author}{\bibfnamefont{K.}~\bibnamefont{Ohmori}},
  \bibinfo{author}{\bibfnamefont{K.}~\bibnamefont{Roumpedakis}},
  \bibnamefont{and}
  \bibinfo{author}{\bibfnamefont{S.}~\bibnamefont{Seifnashri}},
  \bibinfo{journal}{JHEP} \textbf{\bibinfo{volume}{03}}, \bibinfo{pages}{103}
  (\bibinfo{year}{2021}), \eprint{2008.07567}.

\bibitem[{\citenamefont{Koide et~al.}(2022)\citenamefont{Koide, Nagoya, and
  Yamaguchi}}]{Koide:2021zxj}
\bibinfo{author}{\bibfnamefont{M.}~\bibnamefont{Koide}},
  \bibinfo{author}{\bibfnamefont{Y.}~\bibnamefont{Nagoya}}, \bibnamefont{and}
  \bibinfo{author}{\bibfnamefont{S.}~\bibnamefont{Yamaguchi}},
  \bibinfo{journal}{PTEP} \textbf{\bibinfo{volume}{2022}},
  \bibinfo{pages}{013B03} (\bibinfo{year}{2022}), \eprint{2109.05992}.

\bibitem[{\citenamefont{Choi et~al.}(2022{\natexlab{a}})\citenamefont{Choi,
  Cordova, Hsin, Lam, and Shao}}]{Choi:2021kmx}
\bibinfo{author}{\bibfnamefont{Y.}~\bibnamefont{Choi}},
  \bibinfo{author}{\bibfnamefont{C.}~\bibnamefont{Cordova}},
  \bibinfo{author}{\bibfnamefont{P.-S.} \bibnamefont{Hsin}},
  \bibinfo{author}{\bibfnamefont{H.~T.} \bibnamefont{Lam}}, \bibnamefont{and}
  \bibinfo{author}{\bibfnamefont{S.-H.} \bibnamefont{Shao}},
  \bibinfo{journal}{Phys. Rev. D} \textbf{\bibinfo{volume}{105}},
  \bibinfo{pages}{125016} (\bibinfo{year}{2022}{\natexlab{a}}),
  \eprint{2111.01139}.

\bibitem[{\citenamefont{Kaidi et~al.}(2022{\natexlab{a}})\citenamefont{Kaidi,
  Ohmori, and Zheng}}]{Kaidi:2021xfk}
\bibinfo{author}{\bibfnamefont{J.}~\bibnamefont{Kaidi}},
  \bibinfo{author}{\bibfnamefont{K.}~\bibnamefont{Ohmori}}, \bibnamefont{and}
  \bibinfo{author}{\bibfnamefont{Y.}~\bibnamefont{Zheng}},
  \bibinfo{journal}{Phys. Rev. Lett.} \textbf{\bibinfo{volume}{128}},
  \bibinfo{pages}{111601} (\bibinfo{year}{2022}{\natexlab{a}}),
  \eprint{2111.01141}.

\bibitem[{\citenamefont{Hayashi and Tanizaki}(2022)}]{Hayashi:2022fkw}
\bibinfo{author}{\bibfnamefont{Y.}~\bibnamefont{Hayashi}} \bibnamefont{and}
  \bibinfo{author}{\bibfnamefont{Y.}~\bibnamefont{Tanizaki}}
  (\bibinfo{year}{2022}), \eprint{2204.07440}.

\bibitem[{\citenamefont{Choi et~al.}(2022{\natexlab{b}})\citenamefont{Choi,
  Cordova, Hsin, Lam, and Shao}}]{Choi:2022zal}
\bibinfo{author}{\bibfnamefont{Y.}~\bibnamefont{Choi}},
  \bibinfo{author}{\bibfnamefont{C.}~\bibnamefont{Cordova}},
  \bibinfo{author}{\bibfnamefont{P.-S.} \bibnamefont{Hsin}},
  \bibinfo{author}{\bibfnamefont{H.~T.} \bibnamefont{Lam}}, \bibnamefont{and}
  \bibinfo{author}{\bibfnamefont{S.-H.} \bibnamefont{Shao}}
  (\bibinfo{year}{2022}{\natexlab{b}}), \eprint{2204.09025}.

\bibitem[{\citenamefont{Kaidi et~al.}(2022{\natexlab{b}})\citenamefont{Kaidi,
  Zafrir, and Zheng}}]{Kaidi:2022uux}
\bibinfo{author}{\bibfnamefont{J.}~\bibnamefont{Kaidi}},
  \bibinfo{author}{\bibfnamefont{G.}~\bibnamefont{Zafrir}}, \bibnamefont{and}
  \bibinfo{author}{\bibfnamefont{Y.}~\bibnamefont{Zheng}}
  (\bibinfo{year}{2022}{\natexlab{b}}), \eprint{2205.01104}.

\bibitem[{\citenamefont{Roumpedakis et~al.}(2022)\citenamefont{Roumpedakis,
  Seifnashri, and Shao}}]{Roumpedakis:2022aik}
\bibinfo{author}{\bibfnamefont{K.}~\bibnamefont{Roumpedakis}},
  \bibinfo{author}{\bibfnamefont{S.}~\bibnamefont{Seifnashri}},
  \bibnamefont{and} \bibinfo{author}{\bibfnamefont{S.-H.} \bibnamefont{Shao}}
  (\bibinfo{year}{2022}), \eprint{2204.02407}.

\bibitem[{\citenamefont{Bhardwaj et~al.}(2022)\citenamefont{Bhardwaj, Bottini,
  Schafer-Nameki, and Tiwari}}]{Bhardwaj:2022yxj}
\bibinfo{author}{\bibfnamefont{L.}~\bibnamefont{Bhardwaj}},
  \bibinfo{author}{\bibfnamefont{L.}~\bibnamefont{Bottini}},
  \bibinfo{author}{\bibfnamefont{S.}~\bibnamefont{Schafer-Nameki}},
  \bibnamefont{and} \bibinfo{author}{\bibfnamefont{A.}~\bibnamefont{Tiwari}}
  (\bibinfo{year}{2022}), \eprint{2204.06564}.

\bibitem[{\citenamefont{Adler}(1969)}]{Adler:1969gk}
\bibinfo{author}{\bibfnamefont{S.~L.} \bibnamefont{Adler}},
  \bibinfo{journal}{Phys. Rev.} \textbf{\bibinfo{volume}{177}},
  \bibinfo{pages}{2426} (\bibinfo{year}{1969}).

\bibitem[{\citenamefont{Bell and Jackiw}(1969)}]{Bell:1969ts}
\bibinfo{author}{\bibfnamefont{J.~S.} \bibnamefont{Bell}} \bibnamefont{and}
  \bibinfo{author}{\bibfnamefont{R.}~\bibnamefont{Jackiw}},
  \bibinfo{journal}{Nuovo Cim. A} \textbf{\bibinfo{volume}{60}},
  \bibinfo{pages}{47} (\bibinfo{year}{1969}).

\bibitem[{\citenamefont{Hsin et~al.}(2019)\citenamefont{Hsin, Lam, and
  Seiberg}}]{Hsin:2018vcg}
\bibinfo{author}{\bibfnamefont{P.-S.} \bibnamefont{Hsin}},
  \bibinfo{author}{\bibfnamefont{H.~T.} \bibnamefont{Lam}}, \bibnamefont{and}
  \bibinfo{author}{\bibfnamefont{N.}~\bibnamefont{Seiberg}},
  \bibinfo{journal}{SciPost Phys.} \textbf{\bibinfo{volume}{6}},
  \bibinfo{pages}{039} (\bibinfo{year}{2019}), \eprint{1812.04716}.

\bibitem[{\citenamefont{Kapustin and Seiberg}(2014)}]{Kapustin:2014gua}
\bibinfo{author}{\bibfnamefont{A.}~\bibnamefont{Kapustin}} \bibnamefont{and}
  \bibinfo{author}{\bibfnamefont{N.}~\bibnamefont{Seiberg}},
  \bibinfo{journal}{JHEP} \textbf{\bibinfo{volume}{04}}, \bibinfo{pages}{001}
  (\bibinfo{year}{2014}), \eprint{1401.0740}.

\bibitem[{\citenamefont{'t~Hooft}(1980)}]{tHooft:1979rat}
\bibinfo{author}{\bibfnamefont{G.}~\bibnamefont{'t~Hooft}},
  \bibinfo{journal}{NATO Sci. Ser. B} \textbf{\bibinfo{volume}{59}},
  \bibinfo{pages}{135} (\bibinfo{year}{1980}).

\bibitem[{\citenamefont{Fan et~al.}(2021)\citenamefont{Fan, Fraser, Reece, and
  Stout}}]{Fan:2021ntg}
\bibinfo{author}{\bibfnamefont{J.}~\bibnamefont{Fan}},
  \bibinfo{author}{\bibfnamefont{K.}~\bibnamefont{Fraser}},
  \bibinfo{author}{\bibfnamefont{M.}~\bibnamefont{Reece}}, \bibnamefont{and}
  \bibinfo{author}{\bibfnamefont{J.}~\bibnamefont{Stout}},
  \bibinfo{journal}{Phys. Rev. Lett.} \textbf{\bibinfo{volume}{127}},
  \bibinfo{pages}{131602} (\bibinfo{year}{2021}), \eprint{2105.09950}.

\bibitem[{\citenamefont{Choi et~al.}(2022{\natexlab{c}})\citenamefont{Choi,
  Lam, and Shao}}]{Choi:2022jqy}
\bibinfo{author}{\bibfnamefont{Y.}~\bibnamefont{Choi}},
  \bibinfo{author}{\bibfnamefont{H.~T.} \bibnamefont{Lam}}, \bibnamefont{and}
  \bibinfo{author}{\bibfnamefont{S.-H.} \bibnamefont{Shao}}
  (\bibinfo{year}{2022}{\natexlab{c}}), \eprint{2205.05086}.

\bibitem[{\citenamefont{Barkeshli et~al.}(2019)\citenamefont{Barkeshli,
  Bonderson, Cheng, and Wang}}]{Barkeshli:2014cna}
\bibinfo{author}{\bibfnamefont{M.}~\bibnamefont{Barkeshli}},
  \bibinfo{author}{\bibfnamefont{P.}~\bibnamefont{Bonderson}},
  \bibinfo{author}{\bibfnamefont{M.}~\bibnamefont{Cheng}}, \bibnamefont{and}
  \bibinfo{author}{\bibfnamefont{Z.}~\bibnamefont{Wang}},
  \bibinfo{journal}{Phys. Rev. B} \textbf{\bibinfo{volume}{100}},
  \bibinfo{pages}{115147} (\bibinfo{year}{2019}), \eprint{1410.4540}.

\bibitem[{\citenamefont{Gaiotto and Johnson-Freyd}(2019)}]{Gaiotto:2019xmp}
\bibinfo{author}{\bibfnamefont{D.}~\bibnamefont{Gaiotto}} \bibnamefont{and}
  \bibinfo{author}{\bibfnamefont{T.}~\bibnamefont{Johnson-Freyd}}
  (\bibinfo{year}{2019}), \eprint{1905.09566}.

\bibitem[{\citenamefont{Else and Nayak}(2017)}]{Else:2017yqj}
\bibinfo{author}{\bibfnamefont{D.~V.} \bibnamefont{Else}} \bibnamefont{and}
  \bibinfo{author}{\bibfnamefont{C.}~\bibnamefont{Nayak}},
  \bibinfo{journal}{Phys. Rev. B} \textbf{\bibinfo{volume}{96}},
  \bibinfo{pages}{045136} (\bibinfo{year}{2017}), \eprint{1702.02148}.

\bibitem[{\citenamefont{Johnson-Freyd}(2020)}]{Johnson-Freyd:2020twl}
\bibinfo{author}{\bibfnamefont{T.}~\bibnamefont{Johnson-Freyd}}
  (\bibinfo{year}{2020}), \eprint{2011.11165}.

\bibitem[{\citenamefont{Kim}(1979)}]{Kim:1979if}
\bibinfo{author}{\bibfnamefont{J.~E.} \bibnamefont{Kim}},
  \bibinfo{journal}{Phys. Rev. Lett.} \textbf{\bibinfo{volume}{43}},
  \bibinfo{pages}{103} (\bibinfo{year}{1979}).

\bibitem[{\citenamefont{Shifman et~al.}(1980)\citenamefont{Shifman, Vainshtein,
  and Zakharov}}]{Shifman:1979if}
\bibinfo{author}{\bibfnamefont{M.~A.} \bibnamefont{Shifman}},
  \bibinfo{author}{\bibfnamefont{A.~I.} \bibnamefont{Vainshtein}},
  \bibnamefont{and} \bibinfo{author}{\bibfnamefont{V.~I.}
  \bibnamefont{Zakharov}}, \bibinfo{journal}{Nucl. Phys. B}
  \textbf{\bibinfo{volume}{166}}, \bibinfo{pages}{493} (\bibinfo{year}{1980}).

\bibitem[{\citenamefont{Harlow and Ooguri}(2021)}]{Harlow:2018tng}
\bibinfo{author}{\bibfnamefont{D.}~\bibnamefont{Harlow}} \bibnamefont{and}
  \bibinfo{author}{\bibfnamefont{H.}~\bibnamefont{Ooguri}},
  \bibinfo{journal}{Commun. Math. Phys.} \textbf{\bibinfo{volume}{383}},
  \bibinfo{pages}{1669} (\bibinfo{year}{2021}), \eprint{1810.05338}.

\bibitem[{\citenamefont{Coleman and Weinberg}(1973)}]{Coleman:1973jx}
\bibinfo{author}{\bibfnamefont{S.~R.} \bibnamefont{Coleman}} \bibnamefont{and}
  \bibinfo{author}{\bibfnamefont{E.~J.} \bibnamefont{Weinberg}},
  \bibinfo{journal}{Phys. Rev. D} \textbf{\bibinfo{volume}{7}},
  \bibinfo{pages}{1888} (\bibinfo{year}{1973}).

\bibitem[{\citenamefont{Kawasaki et~al.}(2016)\citenamefont{Kawasaki,
  Takahashi, and Yamada}}]{Kawasaki:2015lpf}
\bibinfo{author}{\bibfnamefont{M.}~\bibnamefont{Kawasaki}},
  \bibinfo{author}{\bibfnamefont{F.}~\bibnamefont{Takahashi}},
  \bibnamefont{and} \bibinfo{author}{\bibfnamefont{M.}~\bibnamefont{Yamada}},
  \bibinfo{journal}{Phys. Lett. B} \textbf{\bibinfo{volume}{753}},
  \bibinfo{pages}{677} (\bibinfo{year}{2016}), \eprint{1511.05030}.

\bibitem[{\citenamefont{Nomura et~al.}(2016)\citenamefont{Nomura, Rajendran,
  and Sanches}}]{Nomura:2015xil}
\bibinfo{author}{\bibfnamefont{Y.}~\bibnamefont{Nomura}},
  \bibinfo{author}{\bibfnamefont{S.}~\bibnamefont{Rajendran}},
  \bibnamefont{and} \bibinfo{author}{\bibfnamefont{F.}~\bibnamefont{Sanches}},
  \bibinfo{journal}{Phys. Rev. Lett.} \textbf{\bibinfo{volume}{116}},
  \bibinfo{pages}{141803} (\bibinfo{year}{2016}), \eprint{1511.06347}.

\bibitem[{\citenamefont{Kawasaki et~al.}(2018)\citenamefont{Kawasaki,
  Takahashi, and Yamada}}]{Kawasaki:2017xwt}
\bibinfo{author}{\bibfnamefont{M.}~\bibnamefont{Kawasaki}},
  \bibinfo{author}{\bibfnamefont{F.}~\bibnamefont{Takahashi}},
  \bibnamefont{and} \bibinfo{author}{\bibfnamefont{M.}~\bibnamefont{Yamada}},
  \bibinfo{journal}{JHEP} \textbf{\bibinfo{volume}{01}}, \bibinfo{pages}{053}
  (\bibinfo{year}{2018}), \eprint{1708.06047}.

\bibitem[{\citenamefont{Witten}(1979)}]{Witten:1979ey}
\bibinfo{author}{\bibfnamefont{E.}~\bibnamefont{Witten}},
  \bibinfo{journal}{Phys. Lett. B} \textbf{\bibinfo{volume}{86}},
  \bibinfo{pages}{283} (\bibinfo{year}{1979}).

\bibitem[{\citenamefont{Fischler and Preskill}(1983)}]{Fischler:1983sc}
\bibinfo{author}{\bibfnamefont{W.}~\bibnamefont{Fischler}} \bibnamefont{and}
  \bibinfo{author}{\bibfnamefont{J.}~\bibnamefont{Preskill}},
  \bibinfo{journal}{Phys. Lett. B} \textbf{\bibinfo{volume}{125}},
  \bibinfo{pages}{165} (\bibinfo{year}{1983}).

\bibitem[{\citenamefont{Csaki and Murayama}(1998)}]{Csaki:1998vv}
\bibinfo{author}{\bibfnamefont{C.}~\bibnamefont{Csaki}} \bibnamefont{and}
  \bibinfo{author}{\bibfnamefont{H.}~\bibnamefont{Murayama}},
  \bibinfo{journal}{Nucl. Phys. B} \textbf{\bibinfo{volume}{532}},
  \bibinfo{pages}{498} (\bibinfo{year}{1998}), \eprint{hep-th/9804061}.

\bibitem[{\citenamefont{Witten}(1982)}]{Witten:1982fp}
\bibinfo{author}{\bibfnamefont{E.}~\bibnamefont{Witten}},
  \bibinfo{journal}{Phys. Lett. B} \textbf{\bibinfo{volume}{117}},
  \bibinfo{pages}{324} (\bibinfo{year}{1982}).

\end{thebibliography}

\end{document}